\documentclass[twocolumn,prl,showpacs,amsmath,amstex,amssymb,mathfonts,superscriptaddress]{revtex4-1}

\pdfoutput=1
\usepackage{amsthm,color,amsfonts,graphicx,verbatim}
\usepackage{amsmath}
\usepackage{amssymb}
\usepackage{amsthm}
\usepackage{amsfonts}
\usepackage{listings}
\usepackage{enumerate}
\usepackage{latexsym}
\usepackage{psfrag}
\usepackage{bm}
\usepackage{graphicx}
\usepackage{dsfont}

\usepackage{hyperref}
\hypersetup{
    bookmarks=true,         
    unicode=false,          
    pdftoolbar=true,        
    pdfmenubar=true,        
    pdffitwindow=false,     
    pdfstartview={FitH},    
    pdftitle={Many-body localization in periodically driven systems},    
    pdfauthor={},     
    pdfsubject={},   
    pdfcreator={},   
    pdfproducer={}, 
    pdfkeywords={keyword1} {key2} {key3}, 
    pdfnewwindow=true,      
    colorlinks=true,       
    linkcolor=black,          
    citecolor=blue,        
    filecolor=magenta,      
    urlcolor=blue           
}

\newcommand{\be}{\begin{equation}}
\newcommand{\ee}{\end{equation}}
\newcommand{\bea}{\begin{eqnarray}}
\newcommand{\eea}{\end{eqnarray}}

\newcommand{\la}{\langle}
\newcommand{\ra}{\rangle}

\renewcommand{\phi}{\varphi}
\renewcommand{\epsilon}{\varepsilon}

\begin{document}
\title{Exponentially slow heating in periodically driven many-body systems}

\author{Dmitry A. Abanin}
\affiliation{{Department of Theoretical Physics, University of Geneva, on leave from}}
\affiliation{Perimeter Institute for Theoretical Physics, Waterloo, Canada}
\author{Wojciech De Roeck}
\affiliation{Instituut voor Theoretische Fysica, KU Leuven, Belgium}
\author{Fran\c{c}ois Huveneers}
\affiliation{CEREMADE, Universite Paris-Dauphine, France}

\date{\today}
\begin{abstract}

We derive general bounds on the linear response energy absorption rates of periodically driven many-body systems of spins or fermions on a lattice. We show that for systems with local interactions, energy absorption rate decays exponentially as a function of driving frequency in any number of spatial dimensions. These results imply that topological many-body states in periodically driven systems, although generally metastable, can have very long lifetimes. We discuss applications to other problems, including decay of highly energetic excitations in cold atomic and solid-state systems. 
\end{abstract}

\maketitle

\section{Introduction}
Time-dependent driving recently emerged as a new versatile tool for engineering various quantum states of matter. In non-interacting systems, periodic driving can be used to modify band structures, and in particular to make them topologically non-trivial\cite{Oka09,Kitagawa10,Lindner11,Kitagawa11}. 

Experimentally, periodic driving has been used to realize strong artificial magnetic fields~\cite{exp2}, as well as 2D Bloch bands with non-zero Chern numbers (similar to the Haldane model) in systems of cold atoms in optical lattices~\cite{exp1,exp3}. Theoretically, natural extensions of these ideas to topologically non-trivial Floquet {\it many-body} states have been proposed. In particular, it was suggested that topological Floquet bands can host fractional Chern insulators~\cite{Neupert14}, as well as symmetry-protected topological states~\cite{Santos15}. 

Topological order and more generally quantum order are usually associated with ground states and low temperatures, where they are protected by a finite excitation gap. However, periodic driving breaks energy conservation, making the very concept of the ground state meaningless. It was argued~\cite{Alessio14,Lazarides14,Ponte14} that under driving, generic many-body systems that obey the eigenstate thermalization hypothesis~\cite{deutsch,srednicki,Rigol08}, eventually heat to up to an infinite-temperature, featureless state. Thus, "Floquet topological insulators" are generally metastable. It is important to understand their lifetimes, and use the theoretical understanding to design experiments in which the Floquet many-body states would be long-lived. 

In this paper, we derive general results regarding the heating of periodically driven many-body systems on a lattice. We consider both the cases of local driving (time-dependent perturbation acting only on a few degrees of freedom), and a global driving (driving applied everywhere in the system). The latter setup is relevant to cold atoms experiments~\cite{exp1,exp2,exp3}. Assuming that interactions are local, we prove a general bound for the linear-response heating rates, which indicates that at high driving frequency (much higher than a natural energy scale of the system, e.g., kinetic or interaction energy of one particle), heating is exponentially slow. Fundamentally, this bound follows from the locality of quantum dynamics in systems with local interactions, and, for the case of global driving, relies on the Lieb-Robinson bounds.

\section{Results}

We consider a lattice system of spins or fermions with a local Hamiltonian $H=\sum_{i=1}^N h_i$, subject to a periodic time-dependent perturbation with an operator $O=\sum_i O_i$, which is a sum of one or more local terms $O_i$. {It is understood that $i$ runs over the sites of the lattice and $h_i,O_i$ act on a fixed, finite number of sites around $i$.}
 For simplicity, we focus on the case of harmonic driving with frequency $\omega$ {and strength $g$}: 
\be\label{eq:H}
H(t)=H+g \cos (\omega t) O.
\ee
{and we fix $|| O_{i} || \leq 1$ for concreteness}. 
We assume that the system is initially in a thermal equilibrium, and is described by a density matrix $\rho_\beta$, $\beta=1/T$. 
The energy absorption rate $dE/dt$ is related to  $\sigma(\omega)$, the dissipative part of the linear-response function, by $dE/dt= 2g^2\omega \sigma(\omega)$, {upon averaging over a cycle and up to higher orders in $g$}. The response $\sigma(\omega)$ can be expressed in terms of different-time commutators of operators $O_i$: 
\be\label{eq:sigma}
\sigma(\omega)=\sum_{ij} \sigma_{ij}(\omega), \,\, \sigma_{ij}(\omega)=\frac{1}{2} \int_{-\infty}^{\infty} dt e^{i\omega t} 
\la[O_i (t) , O_j(0)]\ra_\beta , 
\ee
where $\la ...\ra_\beta$ denotes thermal averaging and $O_i(t)$ refers to the Heisenberg dynamics generated by the time-independent $H$. One can show that $\omega \sigma(\omega)$ is positive and symmetric in $\omega$, see e.g.\ \cite{mahanbook,tonglectures}. For concreteness, let us take $\omega\geq 0$ so that $\sigma(\omega)\geq 0$. To be precise,  in finite volume, $\omega \sigma(\omega)$ is a positive distribution rather than a bonafide function and we will need to integrate this distribution over small intervals to state rigorous results: we write $f([\omega_1,\omega_2])\equiv \int_{\omega_1}^{\omega_2} f(\omega)\, d\omega $.

\subsection{Local driving}
First, using the locality of the Hamiltonian, we prove the following bound for individual terms $\sigma_{ij}(\omega)$: 

{\it There is a  $\kappa>0$ and a numerical constant $C$ such that, for any $\delta \omega >0,\omega>0$, 
\be\label{eq:bound1}
|\sigma_{ij}([\omega, \omega+\delta\omega])| \leq C e^{-\kappa |\omega|},
\ee
 uniformly in the total volume.}  
 Note that it is only the dissipative (asymmetric in time) part of the response function that decays exponentially. The reactive (symmetric in time) part contains a principal value integral rather than a $\delta$-function, cfr.\ \eqref{eq:gammaii}, in energy, and it generically decays no faster than $1/\omega$.

We first prove (\ref{eq:bound1}) for $i=j$. It is convenient to write $\sigma_{ii}$ as a sum of contribution of individual eigenstates $|n\ra$ {of $H$}: 
\be\label{eq:sigma_gamma}
\sigma_{ii}(\omega)=\pi \sum_n p_n (\gamma_{ii}^n (\omega)-\gamma_{ii}^n(-\omega)), 
\ee
where $p_n=\frac{e^{-\beta E_n}}{Z}$, $Z={\rm Tr}(e^{-\beta H})$ is the probability that the system is in eigenstate $|n\ra$, and $\gamma_{ii}^n$ is given by
\be\label{eq:gammaii}
\gamma_{ii}^n (\omega)= \sum_{m} |\la m| O_i|n \ra |^2  \delta(E_n-E_m-\omega).
\ee
To estimate this quantity, let us rewrite it as follows, for any $k$: 
\be\label{eq:gamma_modified}
\gamma_{ii}^n (\omega)=\sum_{m} \frac{ |\la m| [[[O_i,H],H],...,H] |n \ra |^2}{\omega^{2k}} \delta(E_n-E_m-\omega), 
\ee
where the r.-h.s. contains $k$ commutators with $H$, and we have used the fact that $E_n-E_m=\omega$.  Next, we use the fact that in local systems, the norm of $[[[O_i,H],H],...,H]$ can be bounded as
\be\label{eq:bound_commutator}
||[[[O_i,H],H],...,H]|| \leq \varepsilon^k k!, 
\ee
where $\varepsilon$  is an energy scale that can be expressed via $||h_i||$ (local norm of the Hamiltonian), the range of $h_j$ and $O_i$, and the coordination number of the lattice. 

Integrating Eq.(\ref{eq:gamma_modified}) over an interval $[\omega,\omega+\delta\omega]$, and using inequality (\ref{eq:bound_commutator}), we obtain: 
\be\label{eq:gamma_estimate}
\gamma_{ii}^n ([\omega, \omega+\delta\omega])\leq \left(  \frac{\varepsilon^k k!}{\omega^k} \right)^2 \leq \left(  \frac{\varepsilon k}{\omega} \right)^{2k}, \,\, \forall k\in \mathbb{N}
\ee
Choosing $k=\frac{ \omega}{\varepsilon e}$, we arrive at the estimate 
\be\label{eq:gamma_estimate2}
\gamma_{ii}^n ([\omega, \omega+\delta\omega])\leq e^{-\kappa \omega}, \,\, \kappa=\frac{2}{\varepsilon e},  
\ee
and therefore, using (\ref{eq:sigma_gamma}) and the fact that $\sum_n p_n=1$, we obtain 
\be\label{eq:sigmaii_estimate}
\sigma_{ii} ([\omega, \omega+\delta\omega])\leq 2\pi e^{-\kappa \omega}. 
\ee

The off-diagonal terms $\sigma_{ij}$, $i\neq j$, can be bounded by diagonal ones using positive-definiteness of $\omega\sigma_{ij}$ and the Cauchy-Schwartz inequality:
\be\label{eq:estimate_off-diag}
|\sigma_{ij}(\omega)|\leq \frac{1}{\sqrt{2}}\left( \sigma_{ii}(\omega)+\sigma_{jj}(\omega) \right). 
\ee
Thus, just like diagonal terms, off-diagonal terms decay exponentially at large frequency, concluding the proof.

The above result can be immediately applied to the case of local driving, when the operator $O$ acts only on a finite number of $n$ lattice sites (i.e.\  $O_i$ on other sites are taken to be zero), while the system size $N$ is taken to infinity. For such a setup, the energy absorption rate will be smaller than $\sim n^2 e^{-\kappa |\omega|}$, where $n$ is the number of lattice sites affected by the periodic driving. 

The result (\ref{eq:bound1}) has a clear intuitive meaning for two particular cases. The first example is a gapped many-body system at zero temperature, e.g., a gapped spin chain with a maximum energy per spin of order $J$. The existence of the gap implies exponential decay of correlations with a typical length scale $\xi$. When the spin chain is subject to a periodic drive, in order to absorb a large energy $\omega$, a local operator $O_i$ must flip a large number $n\sim \omega/J$ spins. It is clear that at large $n\gg \xi$ the amplitude of such a process should be exponentially small. 
Another example is provided by a system with weak interactions, e.g., a weakly interacting Fermi-gas. Then, absorbing a large energy $\omega$ requires a creation of $n\sim \omega/J$ particle-hole pairs ($J$ in this case denotes a maximum energy of one electron-hole pair) -- a process which will be suppressed as $V^n$, where $V$ is the interaction strength. 
Our analysis shows that the energy absorption rate remains exponentially small much more generally: the bound applies to systems without gaps and also to systems with strong interactions. 

\subsection{Global driving}
Next, we consider the case of global driving, when $O_i\neq 0$ for all lattice sites. This setup is generic and relevant, in particular, to all experiments in which driving is used to create topological states. In this case, we prove a different bound: 

{\it There is a  $\kappa>0$ and a numerical constant $C$ such that, for any $\delta \omega >0,\omega>0$,
\be\label{eq:bound2}
\left|\sigma([\omega, \omega+\delta\omega])\right|  \leq N C e^{-\kappa |\omega|}.
\ee
with $N$ the total number of spins.} 
It is intuitively clear that the absorption rate in this case should be proportional to $N$. Therefore one cannot simply use the bound (\ref{eq:bound1}), since it would give a heating rate that scales as $N^2$. Thus, we use Lieb-Robinson bounds \cite{nachtergaele} to estimate $\sigma_{ij}$ for remote $i$ and $j$. For two operators $A,B$ with $|| A ||, ||B || \leq 1$ and support in regions $X,Y$, they read: 
\be\label{eq:LR}
||[A(t), B] ||\leq Ce^{-a(r-v_{LR} t)}, \,\, r={\rm dist} (X,Y), 
\ee
with  $v_{LR}$ the Lieb-Robinson velocity and $C$ a numerical constant (below we always use $C$ for numerical constants whose value can change from line to line).  We choose to measure distances in units of the lattice spacing, and hence $a$ is simply a numerical constant as well. {$a,C,v_{LR}$ are determined by the norm and range of $h_i$ and the type of lattice.}
Without loss of generality, we assume $\kappa \omega \gg 1$ (for small values of $\kappa \omega $, the bound can always be satisfied by tuning $C$). We will prove the bound (\ref{eq:bound2}) for $\delta \omega= (\delta\omega)_0$ with some arbitrary but fixed $(\delta\omega)_0 >0$ (we will use in the proof that $(\delta\omega)_0$ is smaller than quantities that diverge with $\omega \to \infty$). 
The bound for arbitrary $\delta\omega$ is then recovered as follows:
For $\delta\omega < (\delta\omega)_0$, we have $\sigma([\omega,\omega + \delta\omega]) \le \sigma([\omega,\omega + (\delta\omega)_0])$, hence the bound.
For $\delta\omega > (\delta\omega)_0$, we dominate $\sigma([\omega,\omega+\delta\omega]) \le \sum_{k\ge 0} \sigma([\omega + k(\delta\omega)_0,\omega + (k+1)(\delta\omega)_0])$ and we apply the result for each term in the sum, so that the result follows by readjusting $C$.

First, we dominate, for $\omega\geq 2(\delta\omega)_0$,
\be\label{eq:full sigma}
\sigma([\omega, \omega+(\delta\omega)_0]) \leq \frac{e}{1-e^{-8}}   \int_{-\infty}^\infty d\omega' \, e^{-\left( \frac{\omega'-\omega}{(\delta \omega)_0}\right) ^2} \sigma(\omega')
\ee
This relies on positivity of $\sigma(\omega\geq 0)$ and the symmetry $\sigma(-\omega)=-\sigma(\omega) $. Then we split $\sigma(\omega)=\sum_{ij}\sigma_{ij}(\omega)$ in \eqref{eq:full sigma} and we recast the resulting integrals in the time domain: 
\begin{widetext}
\be\label{eq:time_integral}
 \int_{-\infty}^\infty d\omega' \, e^{-\left( \frac{\omega'-\omega}{(\delta \omega)_0}\right) ^2} \sigma_{ij}(\omega')=\sqrt{\pi} (\delta\omega)_0 \int_{-\infty}^\infty dt \, e^{-(t/\delta t)^2} e^{-i\omega t} \la [O_i(t), O_j] \ra_\beta, \qquad \delta t=\frac{2}{(\delta\omega)_0}. 
\ee
\end{widetext}
Our strategy is to estimate terms with ${\rm dist}(i,j)\geq r_*$, with $r_*$  large, using Lieb-Robinson bounds, and to bound terms with ${\rm dist}(i,j)<r_*$ using (\ref{eq:bound1}). 
We will choose  $r_*=2\tilde\kappa \omega/a$ with $\tilde\kappa$ being the $\kappa$ featuring in (\ref{eq:bound1}).

Fist, we study a general $(i,j)$-term in (\ref{eq:time_integral}) with ${\rm dist}(i,j)=r \geq r_*$. We break the time-integral in the r.h.s. of (\ref{eq:time_integral}) into an integral over the interval $[-t_c,t_c]$ with $t_c=t_c(r)=r/(2v_{LR})$, and an integral over the rest of the real axis. The former is bounded by LR-bounds \eqref{eq:LR}:
\be\label{eq:tc}
\left| \int_{-t_c}^{t_c} dt \, e^{-(t/\delta t)^2} e^{-i\omega t}\la [O_i(t),O_j] \ra_\beta \right| \leq 2Ct_c e^{-ar/2}. 
\ee
and the latter is bounded using $|| [O_i(t), O_j] ||\leq 2$ (since $||O_i|| \leq 1$): 
\be\label{eq:tc2}
2\left| \int_{t_c}^{\infty} dt \, e^{-(t/\delta t)^2} e^{-i\omega t}\la [O_i(t),O_j] \ra_\beta \right| \leq 2\sqrt{\pi} e^{-(t_c/\delta t)^2} \delta t. 
\ee
Using (\ref{eq:tc},\ref{eq:tc2}) and setting $\delta t< t_c$ (since $(\delta\omega)_0$ is fixed and $\kappa \omega \gg1$), we bound the sum over $(i,j)$ with  $r\geq r_*$ by 
\be\label{eq:LRestimate}
CN r_*^{d-1}\left[  \frac{r_*}{v_{LR}} e^{-ar_*/2} + \frac{v^2_{LR} (\delta t)^3}{r_*} e^{-\left( \frac{r_*}{2v_{LR} \delta t} \right)^2} \right], 
\ee
By increasing $r_*$, the exponent in the second term becomes at least as small as that in the first term and we can bound  \eqref{eq:LRestimate} by 
$$
 CN r_*^{d}  \left(\frac{1}{v_{LR}} \right)e^{-\tilde\kappa \omega}
$$
 This provides a bound for the contribution of remote pairs $|i-j|\geq r_*$ to the r.-h.s. of Eq.(\ref{eq:time_integral}). Multiplying  by $\sqrt{\pi}(\delta\omega)_0$ and using $ (\delta\omega)_0/v_{LR} \leq C$, we bound their contribution to the response function \eqref{eq:full sigma} by  $ CN  r_*^{d} e^{-\tilde\kappa \omega}$.
There are $\sim r_*^d$ remaining terms with ${\rm dist}(i,j)<r_*$. Their contribution to \eqref{eq:full sigma} can be bounded using (\ref{eq:bound1}), which gives $CN  r_*^d e^{-\tilde\kappa\omega}$, as well.  Recalling $r_*=2\tilde\kappa \omega/a$, we get an overall bound (i.e.\ summed over all $r$) of the form $$C N (\tilde\kappa \omega)^d e^{-\tilde\kappa \omega}$$ for the response function. By slightly reducing $\tilde\kappa$ (the new value is called $\kappa$ again) and increasing $C$, we get the  bound (\ref{eq:bound2}).

The necessity of averaging over a frequency window $\delta \omega$ is likely an artefact of our proof. We expect that in the thermodynamic limit, the bounds (\ref{eq:bound1},\ref{eq:bound2}) hold for $\sigma(\omega)$ itself, i.e.\ that the distribution is a bonafide function, but we cannot prove this, see also \cite{kleinmullerlenoble,kleinmuller} for mathematical details and \cite{brupedra} for a polynomial bound at large $\omega$.

\section{Extensions} Let us briefly comment on extensions of our results. 
First, we note that in $d=1$  we can in fact \emph{choose the decay rate $\kappa$ to be arbitrarily large} at the cost of increasing the numerical prefactors $C$ in (\ref{eq:bound1},\ref{eq:bound2}). This is achieved by improving the bound in the r.h.s. of \eqref{eq:bound_commutator} to  $C(\gamma) e^{-\gamma k} k!$, for any $\gamma>0$, as described in \cite{araki,fnaraki}.  The extension for local driving is immediate and for global driving, we then simply choose $r_*$ with a larger $\tilde\kappa$.  Hence in $d=1$, the decay is in fact superexponential, but one should not expect such an improvement to hold in $d\geq 2$ \cite{bouch}, nor for the case of quasilocal, exponentially decaying $O_i,h_i$ (instead of strictly local). 

It is also possible to obtain similar results as ours for some models of lattice bosons, e.g.\ the Bose-Hubbard model at high temperature, but then the exponential decay is weakened to a stretched exponential. The key observations are that $a)$ the bound \eqref{eq:bound_commutator} fails trivially for unbounded $h_i$ and one has to use a weighted norm instead and $b)$ one needs to assume (spatial) decay of correlations for the thermal ensemble $\langle \cdot \rangle_{\beta}$ (provable by cluster expansions at high temperature) as Lieb-Robinson bounds are no longer available, see \cite{highfreqnonprl}.
 
The most important extension \cite{highfreqnonprl}, however, is to go beyond linear response and to allow for general initial states, showing that the phenomenon described here is quite similar to \emph{localization in energy}, except that, presumably, it in general breaks down after a sufficiently long time, \cite{Alessio14,Lazarides14,Ponte14}.

Finally, for completeness, we mention that there are two remarkable cases where the energy localization does not break down for long times: $1)$ driven MBL systems at not too small frequencies, see \cite{Abanin15,Lazarides15,Ponte15} and $2)$ non-interacting fermions, i.e.\ with $h_i, O_i$ containing only linear and quadratic terms in $c,c^\dagger$. In that case, by performing a canonical transformation, we fall back on a one-particle problem where exact dynamical localization is possible \cite{polkovnikovreview}. In the case where $H$ describes non-interacting fermions but $O$ is of order $q>2$ in the fields $c,c^\dagger$, we do not expect genuine localization but still the linear response vanishes exactly:  $\sigma(\omega)=0$ whenever $\omega$ exceeds the bandwidth times $q$, as one sees from \eqref{eq:gammaii}. 

\section{Discussion} The main implication of the above results is that, although many-body states (e.g.\ topological states) in isolated, periodically driven systems are generally metastable, they have a very long life time, if driving frequency is much higher than the natural energy scale of the system. We note that this limit is indeed realized in recent experiments. 

Further, we note that there has been recent interest in dynamical localization in periodically driven many-body systems~\cite{Prosen98,Alessio13,Lazarides15,Ponte15,Abanin15,Lazarides14,Galitski15}. In such studies, numerical simulations are a useful tool. Our results imply that, in order to observe delocalization at high driving frequency, one may have to study the system dynamics at (exponentially) long times.

Our results also can be directly applied to a different class of problems: the decay of a highly energetic excitation into many low-energy excitations. One physical model where such a problem naturally arises and has been studied is the large-$U$ Fermi-Hubbard model, 
\be\label{eq:fermihubbard}
H=J \sum_{\la i j\ra , s=\uparrow,\downarrow} c_{is}^\dagger c_{js} +U\sum _i n_{i\uparrow} n_{i\downarrow}, \,\, U\gg J
\ee
where $\la ij \ra$ denotes neighbouring sites, and $s$ is a spin label. Doublons (doubly occupied sites) have a typical energy $\sim U$ which is much greater than the kinetic energy $J$. At temperatures $T\ll U$, when there are very few doublons in the system, one can ask how quickly doublons decay into particle-hole pairs. This problem has been addressed experimentally~\cite{doublon1} and theoretically~\cite{doublon2} using perturbation theory, and the decay rates were found to be exponentially small in $U/J$. Therefore, in this case our bound appears to be saturated. 

We note that the bound is non-perturbative in the interaction strength of the systems, and we expect it to be useful for other strongly interacting systems where excitations with very different  energy scales are present, e.g., equilibration of Fermi-Fermi cold gases with very different mass parameters, {as well as random spin models with a broad distribution of exchange couplings. For random spin models, an exponential decay of a local spin correlation function at high frequency has been previously obtained using mean-field-type approximations~\cite{Zobov,Kitaev}. Our results are consistent with this earlier study in $d>1$, but show that in one-dimensional systems the decay at high frequency is even faster than exponential}.

\section{Conclusion}
We have proven that the dissipative part of the linear response for local lattice systems decays exponentially at high frequencies. In particular, this means that heating by periodic driving will be exponentially slow. This result provides a foundation for so-called Floquet many-body phases by showing that, though metastable, they will be very long-lived.\\

{\bf Acknowledgements.} 
D.A.\@ acknowledges support by Alfred Sloan Foundation.  
W.D.R.\@ also thanks the DFG (German Research Fund) and the Belgian Interuniversity Attraction Pole (P07/18 Dygest) for financial support and both F.H.\@ and W.D.R.\@ acknowledge the support of the CNRS Inphynity grant.


\begin{thebibliography}{16}
\bibitem{fnaraki}
{This in turn follows by expanding the repeated commutator as a graphical sum of walks on the lattice (assuming for simplicity that $h_i$ have only nearest neighbour terms) and splitting the walks according to $l \leq k$, the size of the support of the walk on the lattice. The number of walks of length $k$ with given $l$ is then bounded by $2^l l^{k-l}$, with $2$ in $2^l$ being the coordination number in $d=1$. Straightforward algebra shows that $\sum_{l=1}^{k}2^l l^{k-l} \leq C(\gamma) e^{-\gamma k} k!$ for any $\gamma$, proving the claim.}
\end{thebibliography}
\end{document}